\documentclass[journal,twocolumn]{IEEEtran}
\usepackage{fancybox}
\usepackage{xurl}
\usepackage{hyperref}
\hypersetup{colorlinks,allcolors=blue}
\usepackage[inline]{enumitem} 
\usepackage{enumerate} 
\usepackage{environ}
\usepackage{cite}
\usepackage{amsmath,amssymb,amsfonts, multirow, eurosym}
\usepackage{algorithmic, comment}
\usepackage{graphicx}
\usepackage{textcomp}
\usepackage{textcomp, booktabs}
\usepackage[english]{babel} 
\usepackage[utf8]{inputenc} 
\usepackage{multicol} 
\usepackage{flafter} 
\usepackage{soul} 
\usepackage{color} 
\usepackage{array} 
\usepackage{tikz} 

\usepackage{subcaption}

\usepackage{todonotes}
\begin{document}

\title{Dependable Connectivity for \\Industrial Wireless Communication Networks}
\author{Nurul Huda Mahmood,~\IEEEmembership{Member,~IEEE}, Onel L. A. L\'opez,~\IEEEmembership{Senior Member,~IEEE}, David Ruiz-Guirola,~\IEEEmembership{Graduate Student Member,~IEEE}, Frank Burkhardt, Mehdi Rasti and Matti Latva-aho,~\IEEEmembership{Fellow,~IEEE}
\thanks{N.H. Mahmmod, O.L.A. L\'opez, D. Ruiz-Guirola, M. Rasti and M. Latva-aho are with 6G Flagship and Centre for Wireless Communications (CWC), University of Oulu, Finland. Corresponding e-mail:\href{mailto:nurulhuda.mahmood@oulu.fi}{nurulhuda.mahmood@oulu.fi}. F. Burkhardt is with Fraunhofer Institute for Integrated Circuits IIS, Erlangen, Germany. This research has been supported by the Research Council of Finland (RCF) through the projects 6G Flagship (Grant 369116) and ReWIN-6G (Grant 357120).}}



\maketitle

\vspace{-20mm}

\begin{abstract}
Dependability -- a system's ability to consistently provide reliable services by ensuring safety and maintainability in the face of internal or external disruptions -- is a fundamental requirement for industrial wireless communication networks (IWCNs). While 5G ultra-reliable low-latency communication (URLLC) addresses some aspects of this challenge, its evolution toward holistic dependability in 6G must encompass reliability, availability, safety, and security. This paper provides a comprehensive framework for dependable IWCNs, bridging theory and practice. We first establish the theoretical foundations of dependability, including outlining its key attributes and presenting analytical tools to study it. Next, we explore practical enablers, such as adaptive multiple access schemes leveraging real-time monitoring and time-sensitive networking to ensure end-to-end determinism. A case study demonstrates how intelligent wake-up protocols improve event detection probability by orders of magnitude compared to conventional duty cycling. Finally, we outline open challenges and future directions for a 6G-driven dependable IWCN.
\end{abstract}



\section{Introduction}
\label{sec:intro}

\IEEEPARstart{I}{ndustry 4.0} or the fourth industrial revolution is striving for manufacturing efficiency and individual product customization through flexible and reconfigurable manufacturing shop floors. The Industrial Internet of Things (IIoT), including industrial communication networks, is a key enabler of Industry 4.0, where machines and devices equipped with sensors and actuators are connected with advanced analytics and computing to improve the performance and productivity of industrial automation processes~\cite{GRN21_tsn}. Although wired communication is still the dominant connectivity method in IIoT, wireless solutions driven by advances in machine-type communications are gaining significant traction due to their flexibility, scalability, ease of maintainability, and cost-effectiveness~\cite{MAL_20_6GMTC}. 

An industrial communication network, i.e., the communication network in an industrial environment such as a manufacturing plant, is divided into operational technology and information technology. Industrial control systems, such as programmable logic controllers and Supervisory Control and Data Acquisition (SCADA) systems, that form the backbone of operational technology devices require a mix of real-time and isochronous real-time (IRT) communication with deterministic quality of service (QoS) guarantees~\cite{MBA+22_IIoT}. IRT communication is typically characterized by hard guarantees on the cycle time and very low ($1\mu$s or lower) jitter. Existing networking technologies for industrial control systems that meet these requirements are mostly wired solutions that are not interoperable with each other or easily extendable to the wireless domain~\cite{berardinelli2018_wirt}. Thus, there is a need to design a unified wireless IIoT solution for Industry 4.0 that can address the diverse levels of QoS requirements for different verticals, including IRT connectivity.  

The introduction of the ultra-reliable low-latency communications (URLLC) service class in the fifth generation (5G) marked the first step towards having a universal wireless standard to meet these needs. However, the goal of a unified industrial wireless communications network (IWCN) is yet to be realized as URLLC solutions are not efficiently scalable~\cite{park2020extreme}. Moreover, reliability and latency, the main key performance indicators (KPI) in URLLC, are not the only or primary metrics of concern given that knowledge of the distribution of failures is relevant for many industrial use cases~\cite{Khosravirad22_commSurvival}. These limitations can be addressed by incorporating the dependability theory framework from system engineering 
into the URLLC design. 

Dependability is the ability to avoid/mitigate service failures that are more frequent and more severe than acceptable. It is an integrating concept beyond just reliability and low latency provision and aims to directly and holistically enable trustworthy integral systems. Ensuring high dependability mandates a departure from the average utility-based approach of conventional radio resource management (RRM) practices to a framework that considers the distribution and tail behavior of reliability, latency, and other performance metrics under the uncertainty arising from the stochastic nature of wireless communication. 

The implications of URLLC evolution towards dependable communication and strategies to guarantee end-to-end (E2E) dependable IWCNs are discussed in~\cite{Khosravirad22_commSurvival}. Reference~\cite{zhou22_TowardDependable} proposed a semantics-integrated model-driven design approach as a holistic solution for ensuring dependable IWCNs, which is a flexible solution to meet the requirements of more agility and strict timing constraints for automation control systems in modern manufacturing scenarios. A practical mechanism to ensure system-level resilience through extremely high dependability in complex systems consisting of multiple distributed sub-systems is proposed in~\cite{Zunino23_adaptiveSeamless}. The proposed scheme uses a redundant and an adaptive protocol for the data and the control plane, respectively, and is shown to maximize dependability while keeping resource consumption to acceptable levels. Similarly, reference~\cite{lopez2023_multefire} proposed MuLTEfire, a framework for integrating multiple cyber-physical systems (CPS) and IIoT devices through wireless connectivity in the factories of the future. On a similar note, a predictive interference management algorithm for URLLC in beyond 5G networks is studied in~\cite{MLA+20_predictive}. 

Although the above references discuss various dependability enablers, a holistic overview of this emerging concept is still missing and pertinent. This paper aims to fill this gap by providing a holistic introduction to dependable wireless communications, specifically in the context of IWCNs. More precisely, we link theoretical dependability with practical design challenges in IWCNs (cf. Section~\ref{sec:example}) and synthesize key dependability attributes and analytical tools (cf. Section~\ref{sec:theory}). Building on this foundation, we subsequently detail various challenges and explore solution directions in the context of multiple access schemes (cf. Section~\ref{sec:dependableMA}) and time sensitive networking (TSN) (cf. Section~\ref{sec:enablers}). Numerical performance evaluation of a proposed dependability assessment methodology that extends beyond traditional connectivity KPIs to consider the overall goal of timely and relevant event detection is also presented as a concrete example of potential dependability enablers. Finally, the article concludes with an outlook towards dependable communications in future 6G and beyond 6G systems. 

\begin{figure*}[t]
    \centering
    \includegraphics[width=0.7\textwidth]{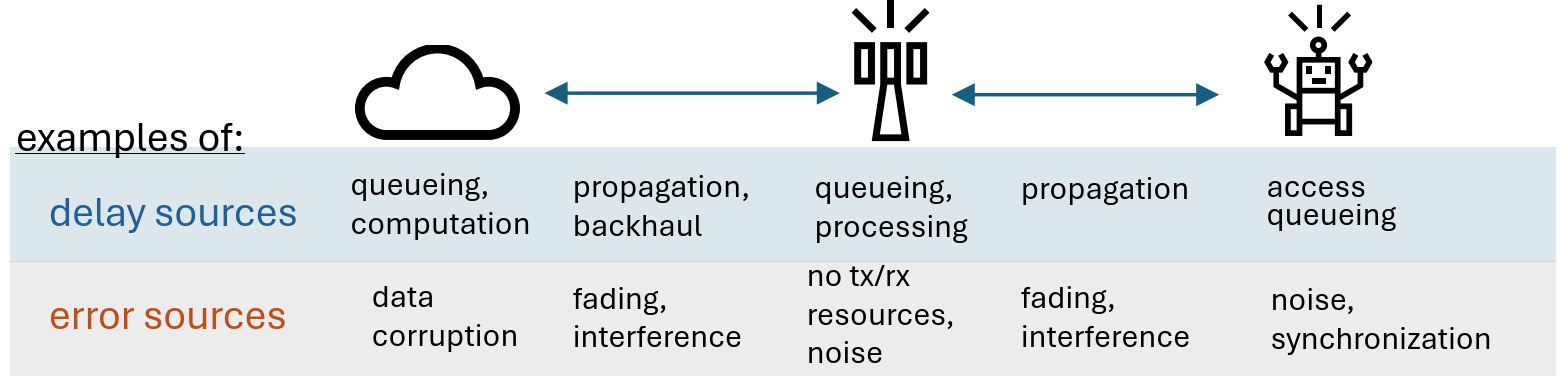}   
    \caption{\small Examples of delay and error sources at the radio access network, core network, and the user-end.}
    \label{fig:delaySources}
\end{figure*}

\section{Why Dependability: Motivation and Challenges} 
\label{sec:example}

New mobile generations have appeared every ten years, each new generation primarily outperforming its predecessor with improved numbers and features. One such example is the URLLC feature introduced to support mission-critical connectivity needs in 5G. In this section, we argue -- through examples from industrial applications -- that while URLLC introduced in 5G addresses some real-world needs, its evolution toward dependable communications is critical for future industrial systems.

\subsection{Motivating Example} 

One of the strongest industrial drivers for URLLC and dependable communication is the rise of industry verticalization and private wireless networks. Unlike public networks, private networks grant enterprises exclusive control over resources, allowing customized service provisioning and strict performance guarantees tailored to specific industrial needs. This shift is particularly significant as industries increasingly rely on CPSs, which are intelligent, interconnected systems where computational and physical processes are deeply integrated. CPSs form the backbone of modern industrial automation, enabling machines, sensors, and controllers to interact in real-time. In these systems, a physical mechanism, such as a robotic arm or an autonomous vehicle, is represented in cyberspace and controlled via computer-based algorithms. CPSs differ from traditional computing systems in that they operate across multiple spatial and temporal scales, exhibit context-dependent behaviors, and dynamically adapt to their environment. As a result, their success relies heavily on precise, predictable, and timely information exchange.

The reliability and dependability notion of timely and successful information delivery in wireless communications is inherently application-dependent. Indeed, the meaning of ``timely'' and ``successful'' varies by context. For example, in human communication, delays below $100$~ms remain imperceptible and relatively frequent faulty information delivery is resolvable/non-critical, while in professional audio applications, latencies above $5$~ms and error rates above $10^{-3}$ become noticeable. In contrast, industrial motion control systems, which dictate the movement of robotic actuators, impose much stricter latency and error rate requirements, as low as $0.5$~ms and $10^{-9}$, respectively. 
Moreover, in such CPSs, the controller periodically sends updates to an actuator at precisely defined intervals. The system is designed to tolerate a certain level of delays and failures in the update process. However, if updates are missing beyond the tolerable limit, the actuator must enter a safe state, halting production to prevent potential damage. This enforced downtime leads to significant productivity losses and, in high-stakes environments, potential safety hazards.

To address these diverse requirements, future networks must adopt a holistic design considering communication, control, computing, and sensing CPS processes. Moreover, application-domain information must be exploited efficiently to predict actual resource requirements, demanding a departure from link-specific URLLC to system dependability analysis/support.

\subsection{Overarching Challenges} 
The multifaceted research challenges and strategic directions essential for realizing the transformative potential of dependable communications include defining precise KPIs and corresponding target values. KPIs and their targets related to performance metrics beyond reliability and latency are yet to be precisely defined or standardized. Meanwhile, application-specific service level agreements (SLA) are crucial for managing dependent applications due to the connectivity and data exchange between service providers, users, and dependent applications. An SLA is an official contract between the service provider and the customer, or between service providers, that defines the priority, quality, and responsibility associated with the provided service. 

Moving beyond KPIs and performance metrics, designing dependable wireless systems and its enablers pose an arduous task due to the fundamental need to identify and accurately characterize the underlying statistical models in which the system operates. These include analyzing the interference statistics, controlling and tracking the channel conditions, modeling the behavior of protocols, and identifying the sources contributing to the delay, errors and jitter~\cite{MBK+23_functional}. The potential delay/jitter and error sources at different network component levels are illustrated in Fig.~\ref{fig:delaySources}. 
Note that each source has its own characteristic statistical behavior that must be understood in order to address them efficiently by adopting proper methodologies and statistical tools~\cite{lopez23_urllcStatistical}. Transmission techniques to mitigate these error sources, such as redundancy and timing guarantees, are discussed in Section~\ref{sec:enablers}.


\section{Dependability Theory}
\label{sec:theory}

It is critical for production lines in the factories of the future to operate smoothly and faultlessly, i.e., every station and component should work as intended. The term \textit{dependability}, i.e., `the ability to perform as and when required', can be used to describe this requirement. 
Dependability as an overarching concept subsumes the attributes of reliability, availability, safety, security, resiliency, etc. The consideration of security brings in concerns for confidentiality, in addition to availability and integrity. Herein, we briefly discuss the main dependability attributes (cf. Section~\ref{sub:dependabilityAttribute}) and explore different tools to analyze them (cf. Section~\ref{sub:dependabilityTools}). 

\subsection{Key Dependability Attributes and Quantities}
\label{sub:dependabilityAttribute}

\begin{figure}[th]
	\centering
	\includegraphics[width=\linewidth]{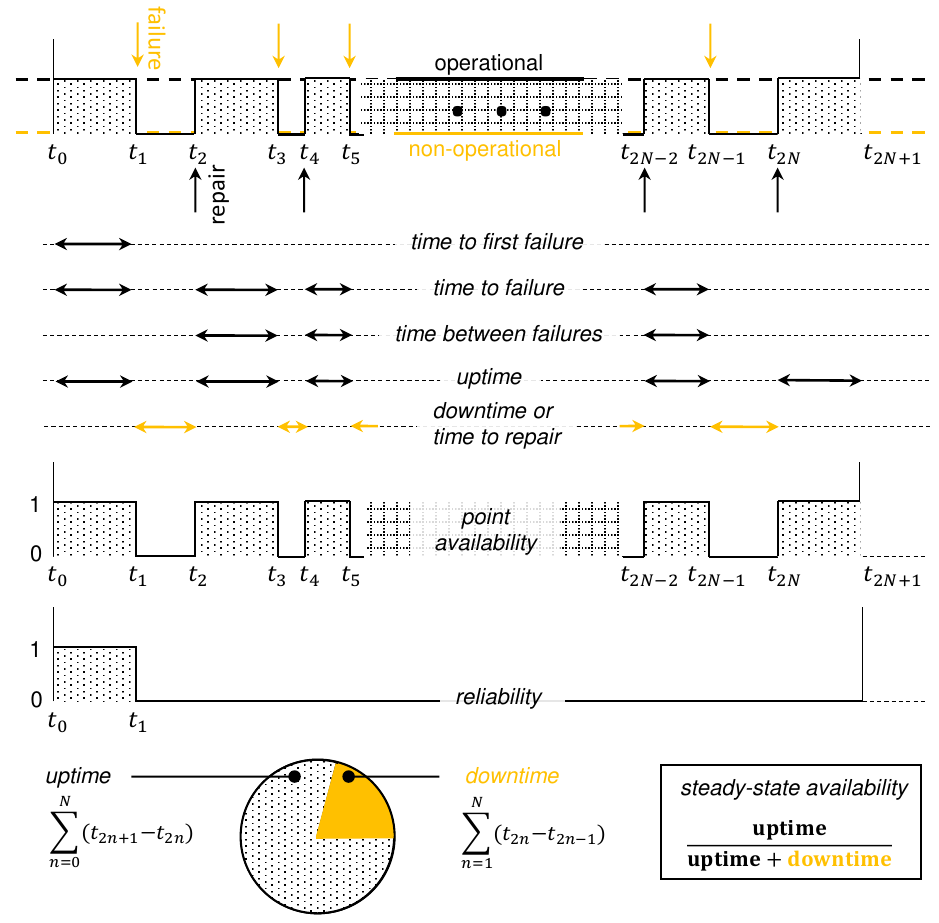}
	\caption{\small Representation of availability, reliability, and maintainability-related metrics.
	}
	\label{Fig_dependability}
\end{figure}

Dependability is an integral concept that cannot be measured by a single metric~\cite{binder2024_dependabilityLookLike}. This section discusses the attributes and quantities that form the guiding principles for dependability analysis and dimensioning. Fig.~\ref{Fig_dependability} depicts and exemplifies some representative quantities, but note that their significance (and that of other metrics) is application-dependent.

\begin{table*}[t]
    \centering
    \begin{tabular}{l p{8.5cm} p{5cm}}
        \toprule
        {Attributes} & Definition & KPIs \\
        \midrule
        \textbf{Availability}~\cite{MBA+22_IIoT} & The \textit{instantaneous availability} at time $t$ is the probability of a system being operational at time $t$ & mean downtime, mean uptime \\
        & \textit{steady-state availability} refers to the long-term probability of being operational at any time & \\
        \textbf{Reliability}~\cite{lopez23_urllcStatistical} & The probability of a system operating without failure during the interval $[0, t]$. Naturally, reliability goes to zero as $t \rightarrow \infty$ & failure distribution, failure rate\\
        \textbf{Safety}~\cite{MBA+22_IIoT} & A system's ability to avoid physical harm, and is a mission-critical applications where human lives are in danger & percentage of time QoS targets are met \\
        \textbf{Security}~\cite{binder2024_dependabilityLookLike} & The capacity of a system to protect itself, the services deployed, and the data exchanged among the system components and users & probability of security breach, attack detection latency, time to mitigation, etc. \\
        \textbf{Resilience}~\cite{mahmood2024resilientbydesign} & Focuses on a system’s ability to recover from and adapt to disruptions, whether caused by faults, failures, security attacks, or environmental challenges & multiple metrics, e.g., mean time to recovery (MTTR), probability of recovery within a deadline\\
         \bottomrule
    \end{tabular}
    \caption{Summary of dependability attributes and related wireless KPIs }
    \label{tab:dependabilitySummary}
\end{table*}

\subsubsection{Availability}

The instantaneous availability of a system at a given time $t$, or point availability, is the probability of being operational at time $t$, while steady-state availability refers to the long-term probability of being operational at any time. The mean downtime and mean uptime are metrics associated with the steady-state availability attribute.
   
\subsubsection{Reliability}
    
Reliability is a time-dependent parameter that refers to the probability of a system operating without failure during the interval $[0, t]$. Reliability goes to zero as $t$ goes to infinity. Two key reliability quantities are the failure distribution and the failure rate. The former characterizes the failure occurrences and the system's operational times, e.g., the time to failure and uptime in memoryless settings. Meanwhile, the failure rate (or Hazard function) $\lambda(t)$ is the conditional probability that a system functioning failure-free until time $t$ also survives an additional small interval $\Delta t$. Notably, the probability of system failure within a short interval $\Delta t$ after operating until time $t$ approximates  $\lambda(t)\Delta t$. In many practical systems where the time to failure is commonly modeled as Weibull distributed, one has that $\lambda(t)\propto t^m, m > -1$. 
    
\subsubsection{Safety}

Safety refers to the ability of a system to avoid harming human life (including properties, animals, etc.). 
It is thus key for mission-critical applications where human lives are in danger, such as autonomous driving and telesurgery. Indeed, faulty conditions such as wrong lane selection in autonomous driving or task offloading failure affecting the information given to the end user and creating distractions in an augmented reality application may bring catastrophic consequences. 
    
\subsubsection{Security}

Security refers to the capacity of a system to protect itself, the services deployed, and the data exchanged among the system components and users. 
Therefore, it encompasses robustness against threats or vulnerabilities such as entity authentication, data confidentiality, outdated data, and data integrity. Many of these features are already considered in wireless systems, but modern dependable systems call for their holistic integration and optimization, especially in the artificial intelligence (AI) era, which can add new security solutions while also making the threat landscape more sophisticated. 

\subsubsection{Resilience}
Resilience focuses on a system's ability to recover from and adapt to disruptions, whether caused by faults, failures, security attacks, or environmental challenges, and is gaining significant attention as a concept on its own~\cite{mahmood2024resilientbydesign}. One key feature here is maintainability, which refers to the ability to repair (or more generally, to successfully modify) a system within a given time. 

{Table~\ref{tab:dependabilitySummary} presents a short comparison summarizing the key dependability attributes and their related wireless KPIs.}

\subsection{Analysis and Support tools}
\label{sub:dependabilityTools}
The previous dependability attributes and quantities become actionable only when one can model and manipulate the elements that give rise to them. This inevitably requires decomposing the system into sub-systems and/or components, {as in \cite{zhou22_TowardDependable}, hereinafter referred to as ``items''. In IWCNs, items may be,
e.g., transmitters, receivers, links/channels, and their sub-components}. These constitute sources of failure and are labeled as repairable or non-repairable according to their life-cycle constraints. The design and analysis of dependable systems require considering all the possible challenges that may occur item-wise and usually involve fault/error/failure prevention, removal, forecasting, and tolerance processes. 

Structure functions simplify analyzing how each item's state affects the system, which can always be decomposed into series and parallel structures for easier evaluation \cite{lopez23_urllcStatistical}. Alongside, fault tree analysis (FTA) and reliability block diagrams (RBD) aid in dependability assessment. FTA is a graphical tool that focuses on system-level risk assessment and mitigation, while RBD illustrates how component reliability impacts system success/failure and is more abstract and easier to handle. 

Both FTA and RBD model logical configurations for reliability analysis but are inherently static and Boolean, opening room for Markovian frameworks, which can accommodate multiple states to capture key temporal and sequential event dependencies{, as in \cite{MLA+20_predictive}}. However, Markov modeling should be avoided when~\cite{lopez23_urllcStatistical}: 
\begin{itemize}
    \item simpler combinatorial methods, e.g., RBD and FTA, suffice,
    \item too many states are needed, 
    \item system behavior is too complex for Markov/semi-Markov models, or 
    \item detailed performance estimates are required.
\end{itemize}


For the latter three cases, simulation or more advanced complex event processing (CEP) tools are usually preferred. CEP enables real-time data analysis for fault detection, tolerance, and proactive measures, and follows two main approaches: time-series modeling and pattern recognition {\cite{lopez23_urllcStatistical,mahmood2024resilientbydesign}}. 
Time-series modeling predicts system behavior from historical data using autoregressive models, Kalman filters, and machine learning (ML) techniques such as recurrent neural networks, long short-term memory networks, and transformers. Meanwhile, pattern recognition leverages ML tools such as support vector machines, neural networks, and clustering to identify meaningful data patterns, involving feature extraction, model training, and classification to detect anomalies and uncover event correlations. 

ML is an inherent part of modern CEP tools, though challenges remain in accurately quantifying uncertainty. Advanced techniques such as Bayesian inference (e.g., Gaussian processes, neural processes), ensemble methods (e.g., deep ensembles, Monte Carlo dropout), and likelihood-based approaches (e.g., variational inference, mixture density networks) can improve confidence estimation {(cf. \cite{lopez23_urllcStatistical} and references therein)}. Additionally, calibration techniques (e.g., temperature scaling) and out-of-distribution detection (e.g., deep kernel learning) further enhance robustness. Integrating these methods is a promising direction towards making CEP-driven systems more dependable in real-time applications.


\section{Dependability Enablers for Multiple Access and Real-time Connectivity}
\label{sec:enablers}

In the previous section, a general overview of the design challenges of reliable communications and potential analytical tools that can be used to solve them was presented. This section details these in the context of two specific application areas, namely multiple access (MA) and TSN. 

\subsection{Dependable Multiple Access Solutions}
\label{sec:dependableMA}

The shared nature of wireless connectivity allows multiple users to communicate over a common set of resources. Access to the shared wireless channel is usually regulated through MA schemes. Dependable MA is crucial for IWCNs where different devices with diverse traffic patterns operating under strict latency requirements may share the same wireless resources. We limit our discussions to uplink MA, which is inherently more challenging than downlink MA due to its distributed, uncoordinated, and sporadic nature. 

Conventional grant-based access schemes incur significant latency and signaling overhead. Meanwhile, Semi-Persistent Scheduling (SPS) improves access efficiency for (semi-)periodic traffic by reducing control overhead but lacks flexibility for dynamic environments, leading to resource waste. Grant-free and non-orthogonal multiple access, currently supported by 5G, removes grant negotiation, improving scalability for sporadic traffic. However, contention-based variants suffer from collisions, whereas contention-free schemes risk resource under-utilization. 
Thus, next-generation dependable MA solutions must adapt to dynamic topologies, heterogeneous QoS demands, power/computation constraints, and harsh environments, e.g., exploiting real-time network/channel monitoring and forecasting tools. We next present two approaches leveraging the latter and conclude this subsection by discussing potential directions for advanced dependable MA schemes.
\subsubsection{Leveraging real-time network/channel monitoring}
%

Adaptive scheduling guided by real-time network/channel monitoring promises dependable MA enhancements. For instance, in network slicing, slices may be dynamically tailored to specific dependability requirements, guaranteeing efficient support for mission-critical applications. Error control mechanisms must also adapt, e.g., forward error correction should dynamically adjust redundancy levels to balance overhead and correction capability;  hybrid automatic repeat request can tune power, coding, and modulation to minimize retransmissions; and network coding strategies in multi-hop networks can adapt to channel/traffic conditions.

Edge ML can facilitate the above and also channel scheduling, both long-term with SPS and real-time through fast uplink grants, wherein devices with expected transmission needs are preemptively allocated spectrum. 
Note that MA's efficacy here depends heavily on the accuracy of traffic predictions, suiting scenarios with time/space-correlated data transmissions where ML techniques such as multi-armed bandit and deep reinforcement learning algorithms can perform well. 
In general, short/long-term-hybrid (SPS/fast-uplink) scheduling is preferred in practice to accommodate different traffic profiles common in industrial networks.

\begin{figure}[t!]
	\centering
	\includegraphics[width=0.7\linewidth]{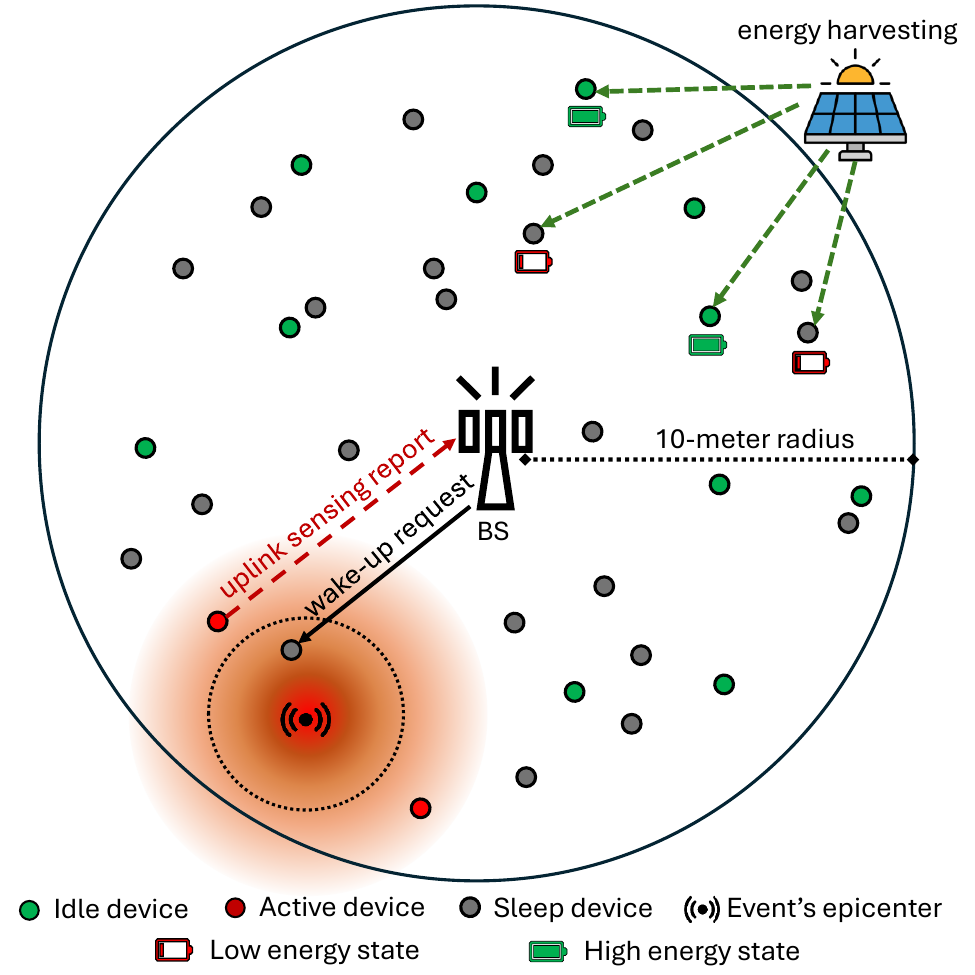}\\
    \caption{$N$ energy-harvesting devices monitoring critical events to report to a BS. Some sleep to conserve energy, while others remain sensing/active for critical event detection. Activation is spatially correlated, with probability decreasing as distance from the event, e.g., a faulty machine vibration epicenter, increases. 
    The BS wakes up and assigns fast grants to devices that may provide additional information about an event, only possible for those at within a maximum distance from the event's epicenter as indicated by the dotted circle.}
   \label{Fig_MA_scenario}
\end{figure}

An example of such a predictive and dynamic MA scheme with intelligent wake-up radios (WuR) is presented next. Consider the setup in Fig.~\ref{Fig_MA_scenario} comprising energy-limited/harvesting devices and a base station (BS). Devices are duty-cycled and their activation depends on proximity to critical events, with active devices transmitting data to the BS upon event detection. To support critical decisions (e.g., machine parameter adjustments or temperature calibration), the BS may request additional data from other sleeping devices equipped with WuR modules that are potentially closer to the event. This can be done by waking up these devices using group (dedicated) wake-up signals (WuS) for multiple (single) devices simultaneously~\cite{Ruiz.2024}. The BS must intelligently manage duty cycling, balancing energy availability and minimizing event reporting latency.   

\begin{figure}[t!]
	\centering
    \includegraphics[width=0.85\linewidth]{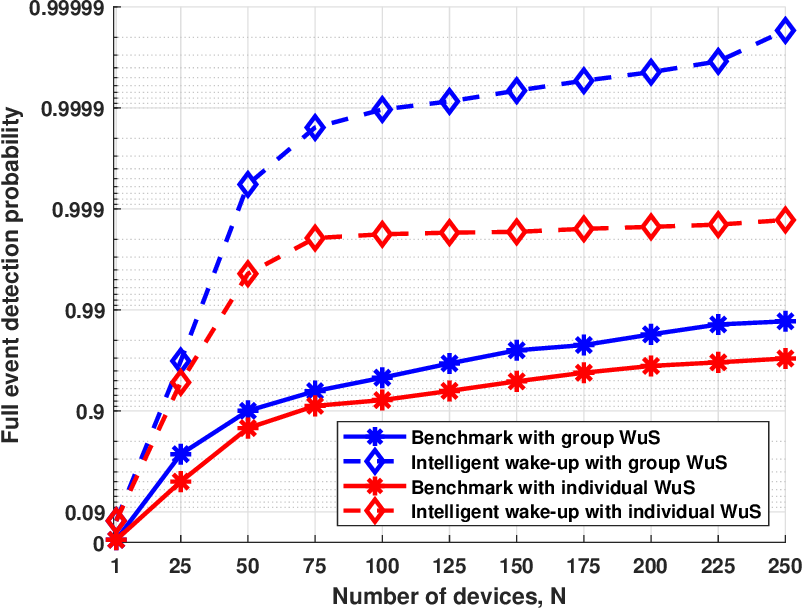}
	\caption{Probability of the BS receiving enough event information within 5 ms vs. the number of devices for an intelligent duty-cycling/wake-up method and a benchmark, and using both group and individual WuS. The benchmark relies solely on spatial correlation, while the intelligent approach exploits the K-nearest neighbor algorithm to dynamically adjust duty cycling and wake-up decisions, incorporating both spatial correlation and battery state predictions. The group-dedicated WuS is formed by $\lceil \sqrt{N} \rceil$ devices, and $e^{-d}$ captures the activation probability of a device at distance $d$ from an event epicenter. The BS initially receives event data within a 1 ms slot, while receiving information from additional devices introduces two extra slots (for the wake-up request and subsequent sensing/transmission).}
   \label{Fig_MA_result}
\end{figure}

Fig.~\ref{Fig_MA_result} compares event detection/characterization capabilities under intelligent BS-managed duty cycling vs. a benchmark using standard duty cycling and spatial correlation-based wake-up. The proposed intelligence improves detection probability by up to three orders of magnitude over the benchmark for the adopted group-dedicated WuS. Improvements are somewhat smaller for device-dedicated WuS, as each request wakes only one device. The gap between the group and the dedicated case widens as the number of devices deployed increases. Note that the performance of group-dedicated WuS depends on the group size: larger groups reduce instantaneous misdetection risks but increase energy resources, potentially impacting long-term device availability. Therefore, optimizing group sizes dynamically and intelligently is a promising research direction.

Our dependability assessment here extends beyond traditional connectivity KPIs, such as reliability and energy depletion probabilities, by considering the overall goal of timely and relevant event detection. Indeed, \textit{even if a device maintains reliable connectivity, its value diminishes if it is not close enough to the event epicenter, potentially compromising long-term system dependability through inefficient energy use.}

\subsubsection{Advanced MA techniques}

Previously discussed approaches, such as adapting physical layer parameters based on real-time feedback on network congestion or interference, necessitate cross-layer design and optimization. Finally, next-generation MA could leverage advances in the following domains to support dependable IWCNs.

\textbf{Secure Multi-Party Computation and Blockchain for MA Coordination:}
Secure multi-party computation and blockchain technologies can enable trust-secured and decentralized coordination among multiple devices, ensuring secure and transparent MA RRM. This enhances network robustness against malicious attacks or misconfigurations that could disrupt communication.

\textbf{Wireless Frame Replication and Elimination for Reliability (FRER):}
FRER, originally developed for TSN (cf. Section~\ref{sec:enablers}), can be extended to wireless MA using network coding. By replicating or coding data packets across multiple paths, wireless FRER may ensure messages reach their destination by mitigating the impact of potential path failures due to high interference or signal blockages.

\textbf{Quantum MA:}
Quantum communication, with the potential for ultra-secure and low-latency data transmissions, could offer revolutionary enhancements to dependability, particularly in scenarios where traditional methods are susceptible to interference or eavesdropping. Research should delve into how quantum entanglement and superposition can apply to MA.


\subsection{TSN as an End-to-End Dependability Enabler}
\label{sec:enablers}
The previous section discussed various enablers for providing access to the shared wireless medium with dependability guarantees. In this section, we briefly present TSN as a solution for ensuring dependable data transmission over the wireless network. IWCNs require dependable connectivity with E2E reliability guarantees. Recent efforts towards integrating 5G into TSN aim to serve this goal. However, this comes with its own set of challenges, especially considering that the delay and jitter variations in wireless networks are significantly larger compared to wired networks. We first briefly present the key TSN features before elaborating on these challenges and potential research directions to address them.

TSN provides a unified layer 1/layer 2 standard for real-time deterministic networks, aiming to guarantee bounded latency by eliminating congestion loss in traffic flows. 
The \textbf{TSN flow concept} introduces different QoS levels by inserting a priority code point tag. \textbf{Flow synchronization} ensures a shared understanding of time across the TSN network through the generic precision time protocol (gPTP), which is required for ensuring IRT communication. \textbf{TSN Flow management}, which can be centralized or distributed, is responsible for admitting or rejecting TSN flows based on the flow QoS requirements and resource availability, and handles the network operations and maintenance function. The handling of different TSN flows with different priority code point tags by the TSN bridges is specified by \textbf{flow control} mechanisms, such as time-aware shaper, frame preemption, and asynchronous traffic shaping. Finally, \textbf{flow integrity} or reliability in TSN is implemented through the FRER, which ensures redundant transmission \& routing, and the per-stream filtering and policing features that offer different mechanisms to protect the intended traffic against bandwidth violation, malfunctioning, congestion, etc.

\begin{figure*}
    \centering
    \includegraphics[width=0.65\textwidth]{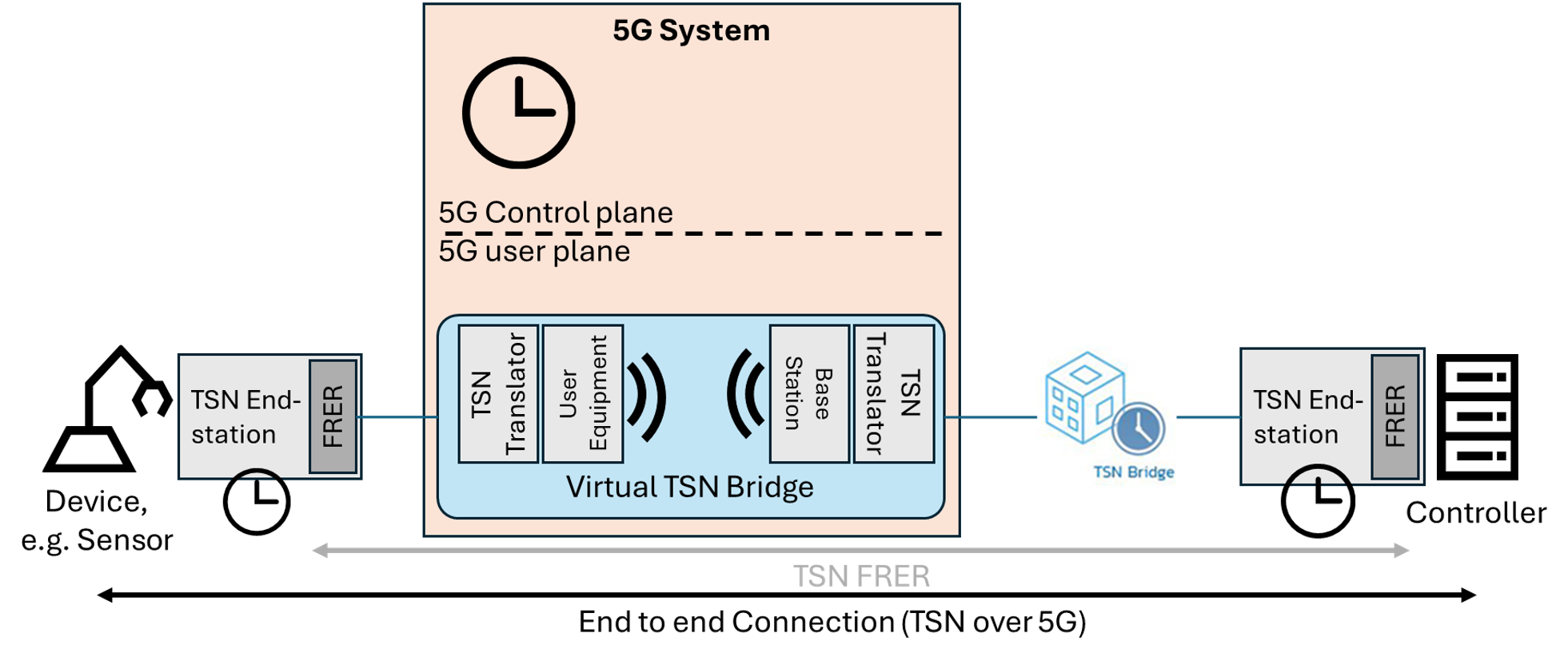}   
    \caption{A simplified illustration of the integration of a 5G network as a logical TSN switch in a TSN network. The TSN translators are deployed at BS and the user equipment side to synchronize with the external TSN network.}
    \label{fig:TSN}
\end{figure*}

As TSN is inherently a wired Ethernet technology built for a static deployment, efforts are being made to support TSN over wireless networks to support deployment flexibility with mobile devices. In terms of 5G, the integration of the 5G network as a logical TSN switch inside a TSN-enabled network is considered the most feasible option~\cite{MBA+22_IIoT}, as illustrated in Fig.~\ref{fig:TSN}. To the outer network, 5G provides the necessary interfaces to transport TSN frames through the TSN translator functionality while using its 5G-specific framework to guarantee services inside the logical switch. Synchronization is maintained by ensuring that the time stamps from the two clocks (the 5G and TSN system clocks) are synced at the TSN/5G bridge at the user equipment side of 5G.

In addition, IEEE aims to include TSN capabilities in WiFi 7 (IEEE 802.11be) to support low latency and ultra-reliability in license-exempt spectrum bands. 
Time synchronization is ensured by operating the gPTP over WiFi, taking the wireless link's asymmetric delay into account, and introducing a novel synchronization method with sub-nanosecond timestamp resolution. The TSN flow concept and control are implemented through traffic prioritization, frame preemption, admission control, and trigger-based scheduled operation. Flow integrity for high reliability and low latency will be ensured through multi-link operation, having two active links operating at different frequencies, and through coordinated transmission across multiple access points. 

Although the initial foundation of a single open standard for a converged network has been laid through the introduction of TSN and TSN-over-wireless, there are still some open questions on how they can operate efficiently. 
Some of these research challenges are briefly described below. 

\textbf{Open Research Questions}

Achieving \textit{high synchronization} accuracy requires frequent control information exchange, resulting in high overhead. The challenge is further exacerbated in wireless networks due to interference, access delays, retransmissions, and mobility issues. High-precision (nanosecond-level) time stamps and accurate calibration of the synchronization signals considering the wireless propagation characteristics are potential solutions towards this end. 
Wireless propagation also introduces significantly larger variations in the packet delay compared to wired networks, resulting in QoS impairments, increased jitter and scalability issues. Candidate techniques to address this include modifying the user scheduling algorithm to shrink the delay variation, i.e., reducing the delay uncertainty, and accounting for the packet delay characteristics of wireless networks in TSN scheduling.

In terms of the \textit{flow control}, the high computational complexity of conventional mechanisms is a major challenge, especially in larger networks. This can be addressed by restricting the solution space by grouping the available flows and restricting the search space of the potential flow route by utilizing contextual flow information. 
QoS guarantees of flows are ensured by reserving E2E resources across the network, but supporting this in a highly scalable and efficient manner is rather challenging. In particular, packet duplication in FRER may induce packet bursts due to the variation in packet delays, thereby degrading the reliability. It also incurs additional delays and congestion (especially in high-traffic scenarios) due to the FRER process. Thus, it is necessary to optimize flow integrity mechanisms considering the network conditions. 

Lastly, simultaneously supporting high reliability that requires a deterministic configuration and a highly flexible network that can meet the \textit{mass customization} demands of Industry 4.0 is another major challenge. This is even more relevant for wireless networks, which must address the vulnerabilities introduced by random fluctuations of the wireless channel, dynamism arising from user mobility and multi-user interference of the shared wireless media. To this end, an in-depth analysis of TSN networks is necessary considering the specific characteristics of wireless networks.


\section{Conclusions}
\label{sec:conc}

The URLLC service class in 5G was introduced to meet the demands of mission-critical applications that require high reliability and low latency to support applications in Industry 4.0 and other verticals where timely and reliable data transmission is crucial. Although this marks an important step towards enabling dependable wireless connectivity, URLLC failed to fully meet dependability needs due to its limited KPIs, scalability constraints, and inability to consider the diverse needs of different applications. To address these limitations, URLLC is expected to evolve towards dependable communications in the 6G era. 

This paper provided a holistic introduction to dependable connectivity for IWCNs, covering both the theoretical aspect and its practical implications. A motivating example of an industrial application requiring dependable connectivity was first presented. We then comprehensively discussed the concept of dependability by introducing different factors influencing it, its theoretical foundation, key attributes $\&$ performance metrics, and various analytical tools to analyze it. Finally, concrete examples of dependability enablers in the context of designing MA and robust data transmission schemes were demonstrated, followed by a highlight on the corresponding open research questions.




\begin{thebibliography}{10}
\providecommand{\url}[1]{#1}
\csname url@samestyle\endcsname
\providecommand{\newblock}{\relax}
\providecommand{\bibinfo}[2]{#2}
\providecommand{\BIBentrySTDinterwordspacing}{\spaceskip=0pt\relax}
\providecommand{\BIBentryALTinterwordstretchfactor}{4}
\providecommand{\BIBentryALTinterwordspacing}{\spaceskip=\fontdimen2\font plus
\BIBentryALTinterwordstretchfactor\fontdimen3\font minus \fontdimen4\font\relax}
\providecommand{\BIBforeignlanguage}[2]{{%
\expandafter\ifx\csname l@#1\endcsname\relax
\typeout{** WARNING: IEEEtran.bst: No hyphenation pattern has been}%
\typeout{** loaded for the language `#1'. Using the pattern for}%
\typeout{** the default language instead.}%
\else
\language=\csname l@#1\endcsname
\fi
#2}}
\providecommand{\BIBdecl}{\relax}
\BIBdecl

\bibitem{GRN21_tsn}
D.~Ginth{\"o}r, Ren{\'e}, N.~Nayak, and J.~von Hoyningen-Huene, ``Time-sensitive networking for industrial control networks,'' in \emph{Wireless Networks and Industrial IoT: Applications, Challenges and Enablers}, N.~H. Mahmood, N.~Marchenko, M.~Gidlund, and P.~Popovski, Eds.\hskip 1em plus 0.5em minus 0.4em\relax Zurich, Switzerland: Springer, 2021, pp. 39--54.

\bibitem{MAL_20_6GMTC}
N.~H. Mahmood \emph{et~al.}, ``Six key features of machine type communication in {6G},'' in \emph{Proc. 2nd 6G Wireless Summit}, Levi, Finland, Mar. 2020.

\bibitem{MBA+22_IIoT}
A.~Mahmood \emph{et~al.}, ``Industrial {IoT in 5G}-and-beyond networks: Vision, architecture, and design trends,'' \emph{IEEE Transactions on Industrial Informatics}, vol.~18, no.~6, pp. 4122--4137, Jun. 2022.

\bibitem{berardinelli2018_wirt}
G.~Berardinelli, N.~H. Mahmood, I.~Rodriguez, and P.~E. Mogensen, ``Beyond {5G} wireless {IRT} for {Industry} 4.0: Design principles and spectrum aspects,'' in \emph{Proc. IEEE Globecom Workshops}, Abu Dhabi, UAE, Dec. 2018.

\bibitem{park2020extreme}
J.~Park \emph{et~al.}, ``Extreme ultra-reliable and low-latency communication,'' \emph{Nature Electronics}, vol.~5, pp. 133--141, Mar. 2022.

\bibitem{Khosravirad22_commSurvival}
S.~R. Khosravirad \emph{et~al.}, ``Communications survival strategies for industrial wireless control,'' \emph{IEEE Network}, vol.~36, no.~2, pp. 66--72, May 2022.

\bibitem{zhou22_TowardDependable}
N.~Zhou \emph{et~al.}, ``Toward dependable model-driven design of low-level industrial automation control systems,'' \emph{IEEE Transactions on Automation Science and Engineering}, vol.~19, no.~1, pp. 425--440, Jan. 2022.

\bibitem{Zunino23_adaptiveSeamless}
C.~Zunino, G.~Cena, S.~Scanzio, and A.~Valenzano, ``Adaptive seamless redundancy to achieve highly-dependable mqtt communication,'' \emph{IEEE Transactions on Industrial Informatics}, pp. 1--11, 2023.

\bibitem{lopez2023_multefire}
M.~López \emph{et~al.}, ``Towards the {5G}-enabled factories of the future,'' in \emph{2023 IEEE 21st International Conference on Industrial Informatics (INDIN)}, Lemgo, Germany, 2023, pp. 1--8.

\bibitem{MLA+20_predictive}
N.~H. {Mahmood}, O.~A. {Lopez}, H.~{Alves}, and M.~{Latva-aho}, ``A predictive interference management algorithm for {URLLC} in beyond {5G} networks,'' \emph{IEEE Communications Letters}, vol.~25, no.~3, pp. 995--999, Mar. 2021.

\bibitem{MBK+23_functional}
N.~H. Mahmood \emph{et~al.}, ``A functional architecture for {6G} special-purpose industrial {IoT} networks,'' \emph{IEEE Transactions on Industrial Informatics}, vol.~19, no.~3, pp. 2530--2540, 2023.

\bibitem{lopez23_urllcStatistical}
O.~L.~A. López \emph{et~al.}, ``Statistical tools and methodologies for ultrareliable low-latency communication—a tutorial,'' \emph{Proceedings of the IEEE}, vol. 111, no.~11, pp. 1502--1543, 2023.

\bibitem{binder2024_dependabilityLookLike}
R.~V. Binder, ``What does dependability look like?'' \emph{IEEE Reliability Magazine}, vol.~1, no.~2, pp. 12--16, 2024.

\bibitem{mahmood2024resilientbydesign}
\BIBentryALTinterwordspacing
N.~H. Mahmood \emph{et~al.}, ``Resilient-by-design: A resiliency framework for future wireless networks,'' 2024. [Online]. Available: \url{https://arxiv.org/abs/2410.23203}
\BIBentrySTDinterwordspacing

\bibitem{Ruiz.2024}
D.~E. Ru{\'\i}z-Guirola \emph{et~al.}, ``Intelligent duty cycling management and wake-up for energy harvesting {IoT} networks with correlated activity,'' in \emph{Proceedings of the Asilomar Conference on Signals, Systems, and Computers}, California, US, Oct. 2024.

\end{thebibliography}

\end{document}